\documentclass[a4paper,10pt,twocolumn,superscriptaddress,aps,pre]{revtex4-1}
\usepackage[latin1]{inputenc}
\usepackage{amsmath}
\usepackage{amsfonts}
\usepackage{amssymb}
\usepackage{graphicx}
\usepackage{color}
\usepackage{soul}

\graphicspath{{figures/}}

\begin{document}
\title{Perturbing the Shortest Path on a Critical Directed Square Lattice}
\date{\today}

\author{Fabian Hillebrand}
\affiliation{ETH Z\"urich, Computational Physics for Engineering Materials, Institute for Building Materials, Wolfgang-Pauli-Str. 27, HIT, CH-8093 Z\"urich (Switzerland)}
\author{Mirko Lukovi\'c}
\affiliation{ETH Z\"urich, Computational Physics for Engineering Materials, Institute for Building Materials, Wolfgang-Pauli-Str. 27, HIT, CH-8093 Z\"urich (Switzerland)}
\author{Hans J. Herrmann}
\affiliation{ETH Z\"urich, Computational Physics for Engineering Materials, Institute for Building Materials, Wolfgang-Pauli-Str. 27, HIT, CH-8093 Z\"urich (Switzerland)}
\affiliation{Departamento de F\'isica, Universidade do Cear\'a, 60451-970 Fortaleza (Brazil)}
\affiliation{PMMH, ESPCI, 7 quai St Bernard, 75005 Paris (France)}
\begin{abstract}
	We investigate the behaviour of the shortest path on a directed two-dimensional square lattice for bond percolation at the critical probability~$p_c$ . We observe that flipping an edge lying on the shortest path has a non-local effect in the form of power-law distributions for both the differences in shortest path lengths and for the minimal enclosed areas. Using maximum likelihood estimation and extrapolation we find the exponents $\alpha = 1.36 \pm 0.01$ for the path length differences and $\beta = 1.186 \pm 0.001$ for the enclosed areas.
\end{abstract}

\maketitle

\section{Introduction}
The well established bond percolation model can be generalized on a directed lattice as was done by Redner with the resistor-diode network, which was used implicitly much earlier in the work of Broadbent and Hammersley \cite{redner1981,redner1982a,redner1982b,broadbent1957,noronha2018,janssen2000}.

For bond percolation the shortest path $l$ is defined as the path connecting two opposite sides of a lattice with the least amount of steps. The length $l$ is also known as the chemical distance \cite{havlin1984}. On a randomly directed lattice the path, in addition, has to respect the direction of each link. Shortest paths of two-dimensional percolating systems have been investigated extensively over the past 40 years and have been found to be fractal at the percolation threshold \cite{grassberger1983,pike1981,herrmann1984,grassberger1985,herrmann1988,zhou2012}.

\begin{figure}[ht]
	\begin{center}
		\includegraphics[scale=0.35]{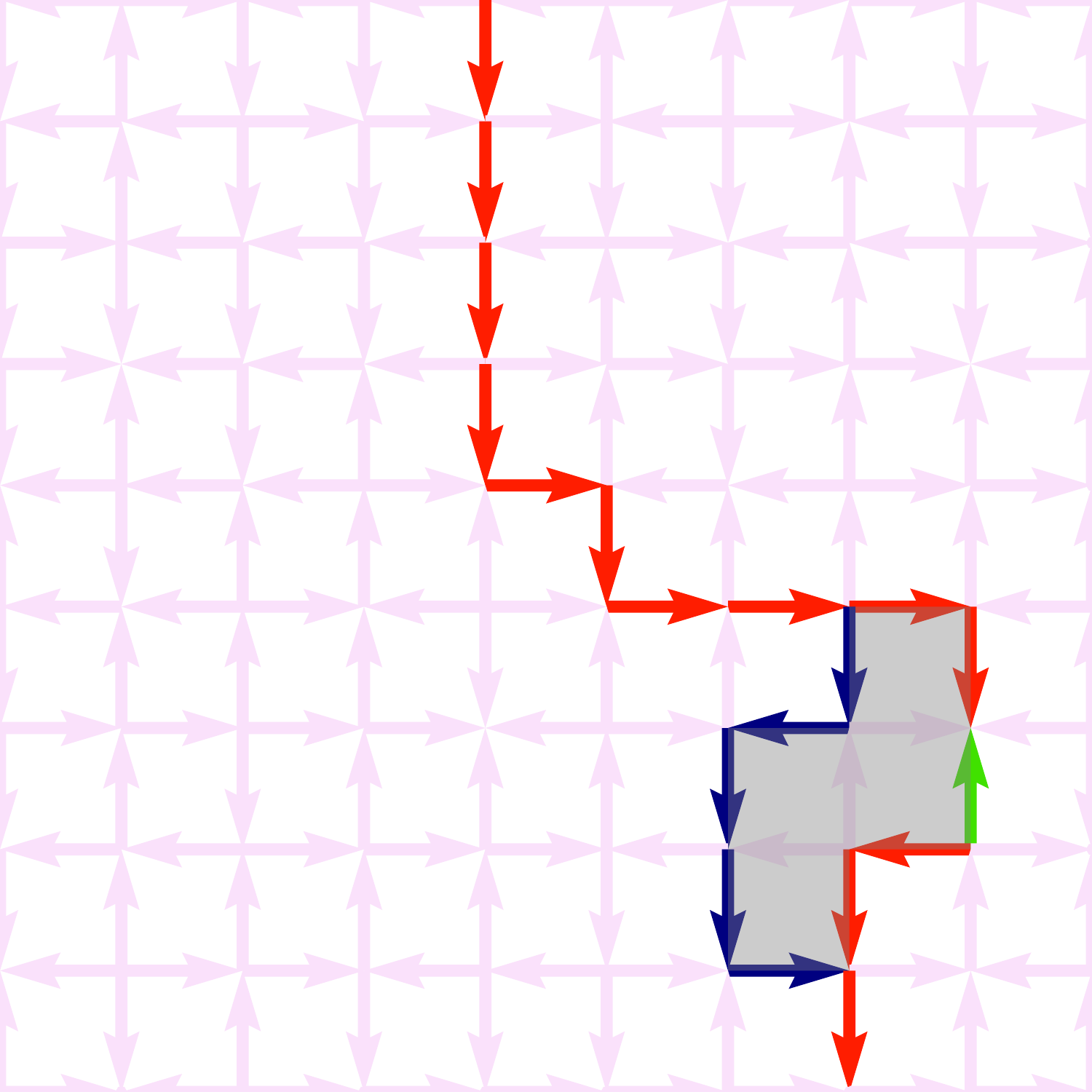}
	\end{center}
	\caption{The original shortest path is in red. After inverting one of the links on the path (green arrow), the shortest path changes. The new shortest path is shown in blue. It is chosen out of all possible shortest paths such that it minimizes the area of the shaded region.}
	\label{fig:shortes_path}
\end{figure}

We focus on directed networks where all the directions are equiprobable, thus making the system critical on the square lattice \cite{redner1981}. We consider a two-dimensional directed square lattice with side length~$L$ and investigate how flipping an edge (i.e.\ inverting the direction of an edge) affects the shortest path length~$l$. An example of such a situation is shown in figure \ref{fig:shortes_path}. When the shortest path is interrupted by inverting a randomly selected bond on the path, a new distinct shortest path emerges. We use numerical simulations to investigate the difference in length between the new and old shortest path and the area enclosed between them.

We use the burning method to determine the shortest path that connects one side (top) of the lattice with the other (bottom) \cite{herrmann1984}. The burning method starts by setting all vertices in the topmost row on 'fire'. Then, in the following time steps, the fire is propagated along the permissible edges to neighbouring sites that are not burnt. At each time step, the newly burnt vertices are labelled with the current time. The procedure stops as soon as the opposite side is reached or earlier in cases where the system does not percolate. Assuming the fire reaches the opposite side, the shortest path length is given by the last time step to be registered. It should be noted that shortest paths on square lattices are usually not unique.

We choose randomly one of the shortest paths determined using the burning method. A randomly selected edge along this path is then flipped and the burning method employed again to find the new shortest path $l_{new}$. The one that encloses the smallest area together with the original shortest path is selected. Choosing the smallest area still leaves potentially two candidates for the shortest path, but they have the same enclosed area and difference in shortest path lengths. By flipping one of the bonds that lie on the shortest path we introduce an external perturbation with the goal of studying how the system, and in particular, the shortest path responds to it. We want to understand whether these local modifications can result in much larger global changes.

\begin{figure}[ht]
	\begin{center}
		\includegraphics[scale=0.35]{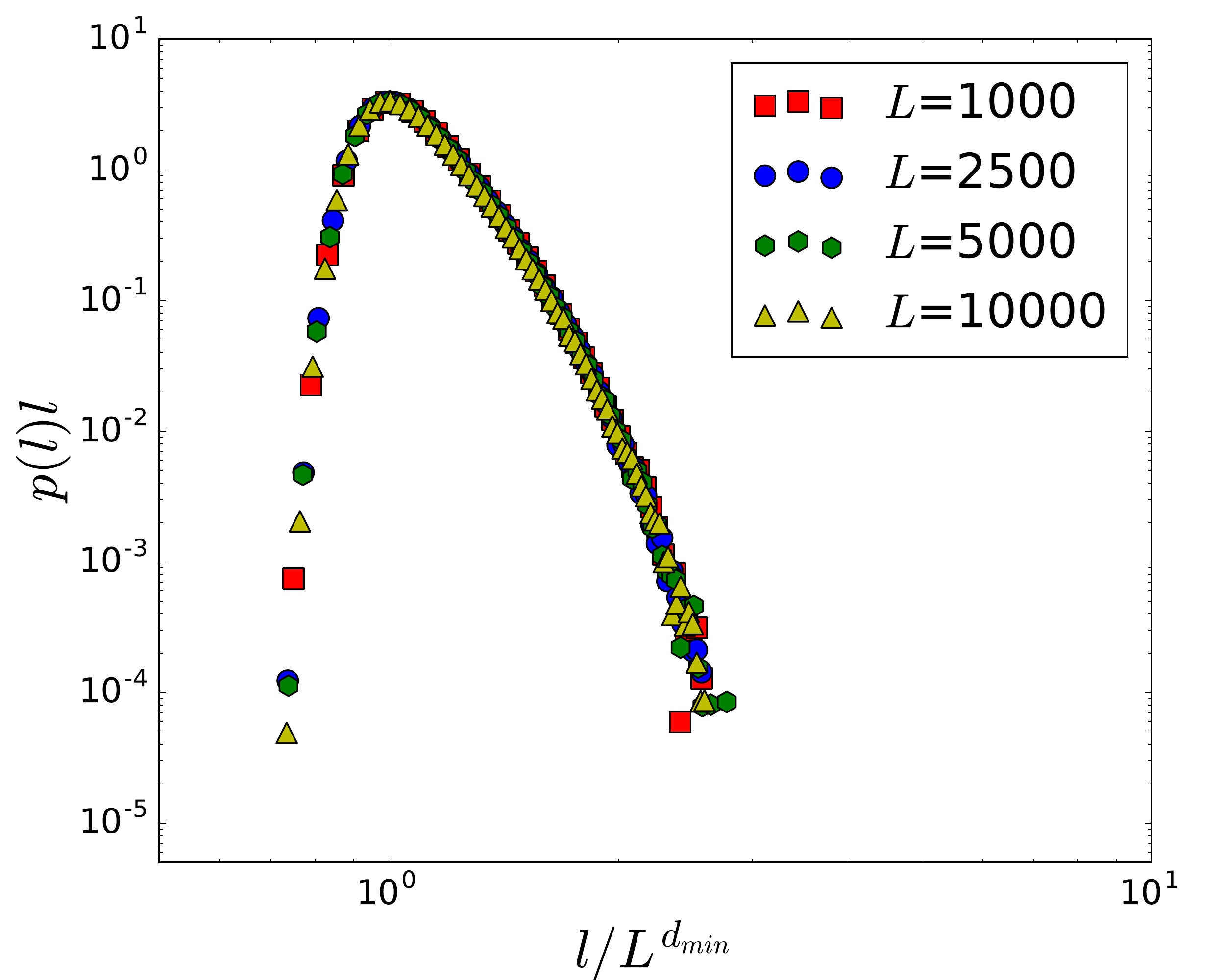}
	\end{center}
	\caption{Data collapse for the distribution of shortest path lengths $p(l)$ using data for different lattice sizes $L$. The exponent $d_{min}=1.13$ is the fractal dimension of the shortest path.}
	\label{fig:shortes_path_distro}
\end{figure}

As a first step, we have a look at the distribution of the unperturbed shortest paths~$p(l,L)$. It turns out that also in the case of directed networks, $p(l,L)$ is compatible with the same scaling law found for bond percolation, \cite{neumann1988,havlin1985,aharony1992,hovi1997}.  From our simulation results we find that
\begin{equation}
	p(l,L)=g(l/L^{d_{min}})/l,
\end{equation}
where $L$ is the size of the square lattice and $d_{min}=1.13$ the fractal dimension of the shortest path \cite{zhou2012,aharony1992}. This comes as no surprise since it has already been conjectured that the model discussed in this manuscript is in the same universality class as standard percolation \cite{janssen2000,zhou2012b}. The data collapse of our results for different systems sizes $L$, and therefore the finite-size scaling, can be observed in figure \ref{fig:shortes_path_distro}. Furthermore, we also confirm the analytical asymptotic results obtained by Hovi and Aharony for the scaling function $g(x)$ (not shown).

\section{Change in shortest path length}

On square lattices the length difference between the old shortest path and the new one is always an integer and it can be zero, even or odd. Two paths sharing the same starting and ending points as shown in figure \ref{fig:shortes_path} can either have a zero or an even difference in length. This is because the length of any cycle on a square lattice is an even number and thus the difference between any two paths that agree on both ends will be either zero or even. An odd value, on the other hand, implies that at least one of the two end points does not coincide. 

The change in length of the shortest path necessarily has to be non-negative. For a new path to be shorter than the previous path, it has to go through the flipped edge because otherwise it would mean that the new path already existed before the change. On the other hand, any new shortest path cannot pass through the flipped edge as illustrated by the following argument. Consider 4 paths $A$, $B$, $C$ and $D$. The paths $A$ and $B$ start from the top and go each to one end of the flipped edge $e$ while $C$ and $D$ start each from one end of the flipped edge $e$ and go to the bottom. Let the path $A + e + C$ denote the previous shortest path. Before flipping, both paths $A + D$ and $B + C$ must be as long or longer than $A + e + C$ by definition. In turn this means that $B$ is longer than $A$ and $D$ is longer than $C$. When the edge $e$ is now flipped, we observe that $B + e + D$ is longer than both $A + D$ and $B + C$. Therefore, the flipped edge will never be crossed in any new shortest path.

\begin{figure}[ht]	
	\begin{center}
		\includegraphics[scale=0.34]{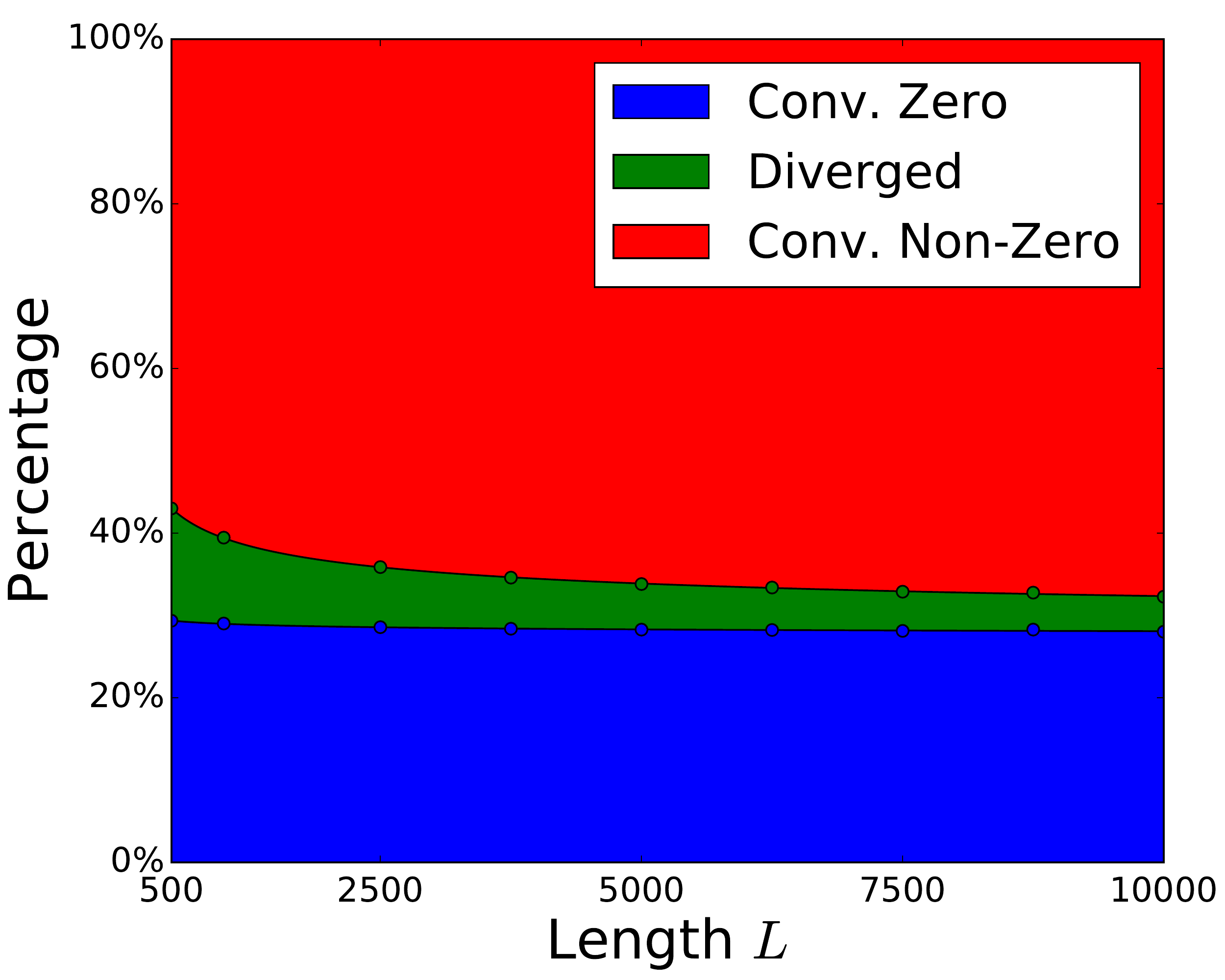}
	\end{center}
	\caption{Ratio of the three possible outcomes for the difference between the old and new shortest path:  convergent paths with zero difference in blue, divergent paths in green and convergent paths with non-zero difference in red.}
	\label{fig:percentage}
\end{figure}

We separate the data obtained from the simulations based on whether the shortest paths agree on both ends and then further based on whether the differences between them is zero or non-zero. If the two shortest paths do not agree on both ends, we say that they are divergent and otherwise we say that they are convergent. To summarize, the data is classified into divergent paths, convergent paths with zero differences and convergent paths with non-zero, i.e.\ even, differences. The ratio of these three categories obtained as a function of system size is shown in figure \ref{fig:percentage}. We extrapolate the curves by fitting $\gamma(L) = \gamma_0 + \gamma_s L^{\gamma_e}$ using non-linear least squares as shown in figure \ref{fig:percentage}.  For $L\rightarrow\infty$ we observe that the samples with convergent paths with zero change in length approach  $26.9\%\pm1.1\%$, those with divergent paths approach $0.0\%\pm0.2\%$ and those with convergent paths with non-zero change in length approach $72.8\%\pm0.5\%$. In comparison, if the divergent paths are ignored explicitly, the percentage for zero-difference paths becomes $27.4\%\pm0.4\%$ and for even differences $72.6\%\pm0.4\%$. We can therefore conclude that the probability of obtaining divergent paths vanishes in the asymptotic limit of large systems sizes.

\begin{figure}[ht]	
	\begin{center}
		\includegraphics[scale=0.35]{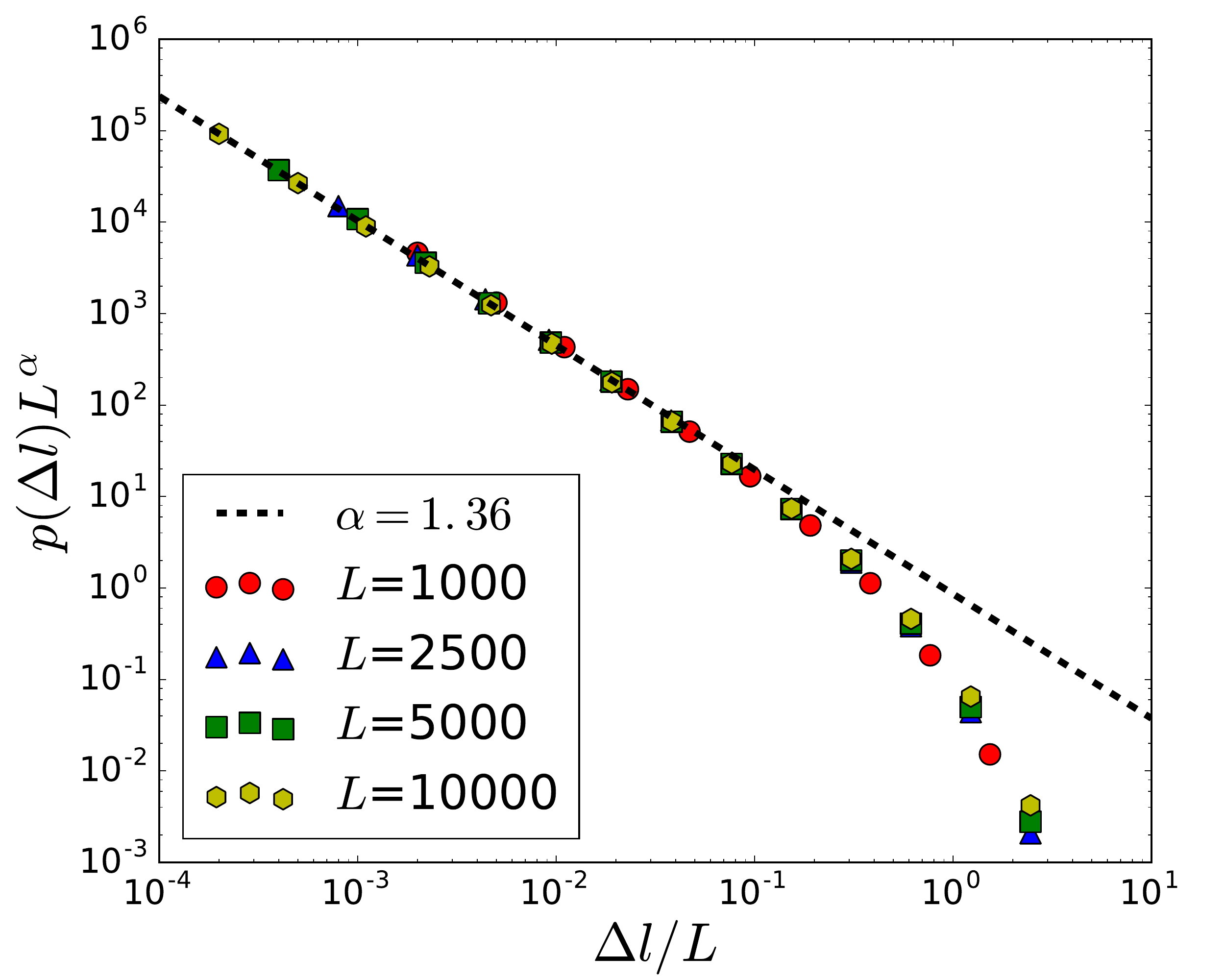}
	\end{center}
	\caption{Data collapse for the distribution of non-zero differences in shortest path lengths $p(\Delta l)$ for convergent paths using data for different lattice sizes $L$. The dashed line represents a power law fit to the data with exponent $\alpha=1.36$.}
	\label{fig:shortes_path_change_distro}
\end{figure}

\begin{figure}[ht]	
	\begin{center}
		\includegraphics[scale=0.35]{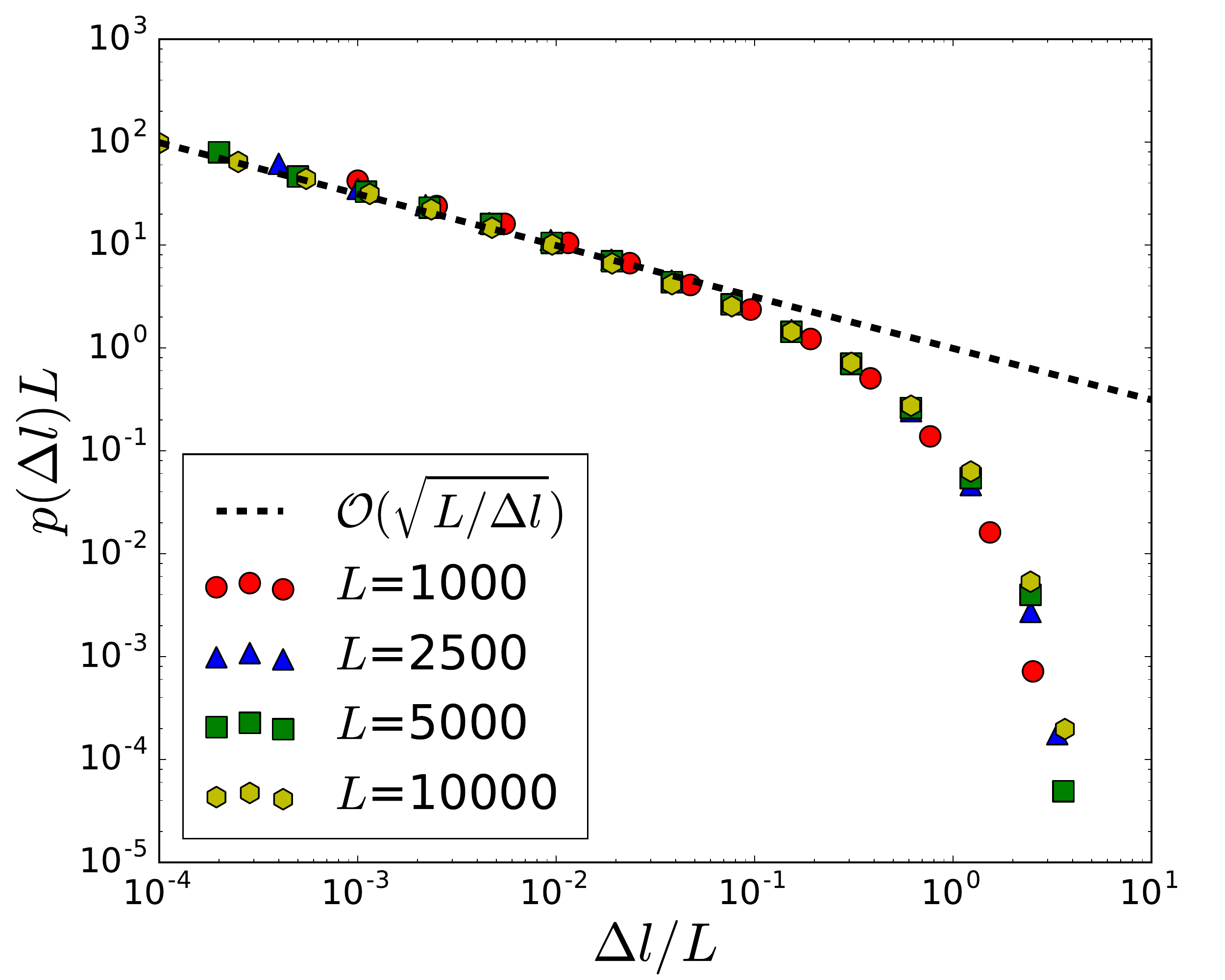}
	\end{center}
	\caption{Data collapse for the distribution of non-zero differences in shortest path lengths $p(\Delta l)$ for divergent paths using data for different lattice sizes $L$. The dashed line represents a power law curve with exponent $\alpha = 0.5$.}
	\label{fig:shortes_path_change_odd_distro}
\end{figure}

If we consider only the convergent paths with non-zero differences, the distribution~$p(\Delta l)$ of the differences in length between old and new shortest paths shows scale-free behaviour with an exponent estimated to be $\alpha = 1.36 \pm 0.01$. Figure \ref{fig:shortes_path_change_distro} shows an excellent data collapse for the distribution~$p(\Delta l)$ that obeys scaling of the form $p(\Delta l)=L^{-\alpha} f(\Delta l / L )$. In the case of divergent paths, again only with non-zero differences, the results are quite different as can be seen in figure \ref{fig:shortes_path_change_odd_distro}. A truncated power law is observed, but with an exponent of approximately $\alpha\simeq 0.5$, which is different from the scaling exponent of unity used to obtain the data collapse. Therefore, when considering only divergent paths, the probability distribution $p(\Delta l,L)$ will tend to zero as $L$ tends to infinity. This result further supports our claim that the fraction of diverging paths decays to zero for $L\rightarrow\infty$.

\section{Area between shortest paths}

\begin{figure}[ht]
\begin{center}
\includegraphics[scale=0.35]{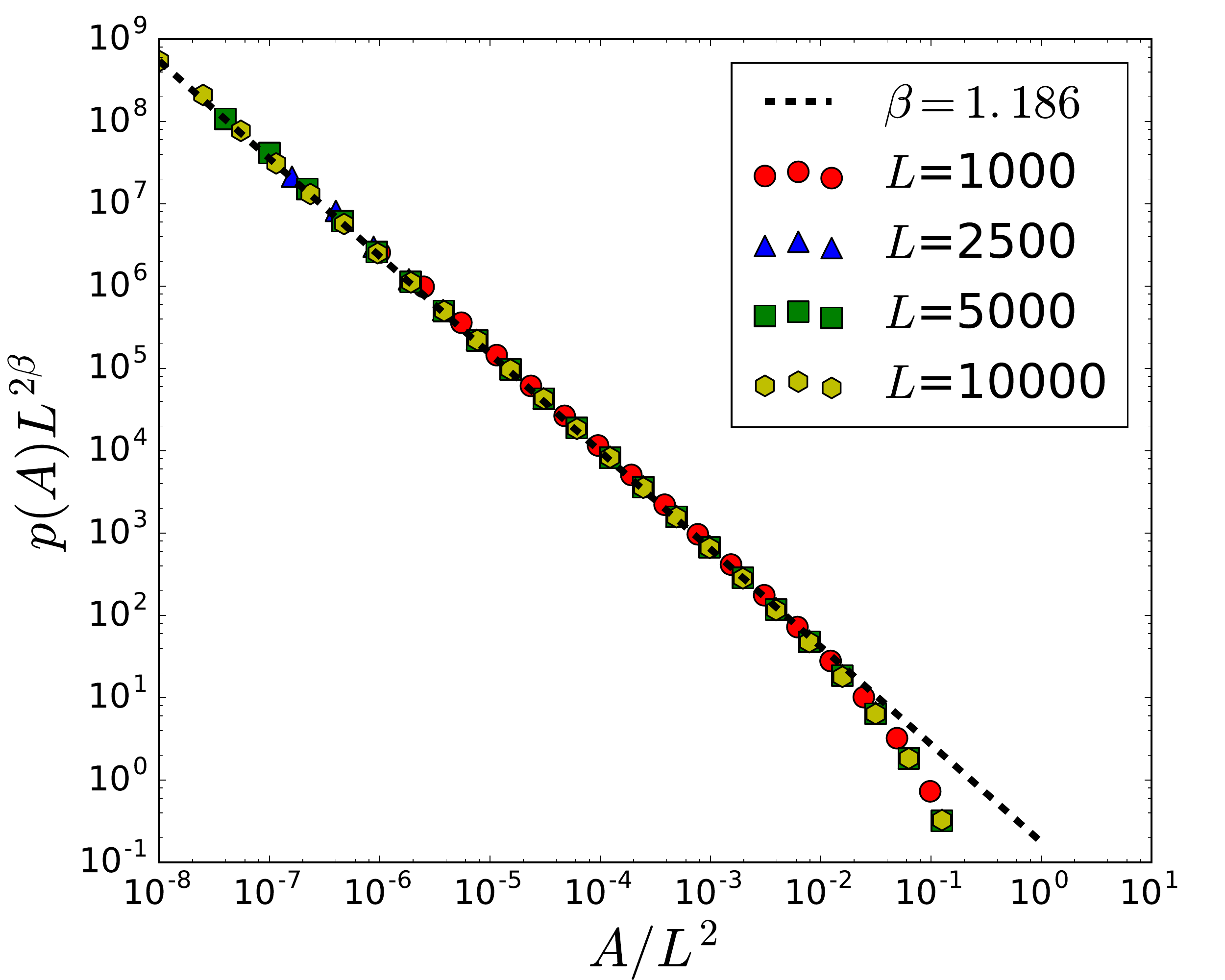}
\end{center}
\caption{Data collapse for the distribution of enclosed areas $p(A)$ using data for different lattice sizes $L$. The dashed line represents a power law fit to the data with exponent $\beta=1.186$.}
\label{fig:area_distro}
\end{figure}

The distribution~$p(A)$ of the size of the minimal area~$A$ enclosed between the old and new shortest paths also shows scale-free behaviour with an exponent estimated to be $\beta = 1.186 \pm 0.001$. Figure \ref{fig:area_distro} shows an excellent data collapse for the distribution $p(A)$ that obeys scaling in the form $p(A)=L^{-2\beta} f(A / L^2 )$. The samples used only contain convergent paths. The exponent $\beta$ agrees within error bars with the exponent $\beta = 1.16 \pm 0.03$ for the size distribution of the areas enclosed by watersheds from landscapes that only differ slightly at one location reported in Ref. \cite{fehr2011} in the case of uncorrelated landscapes with Hurst exponent $H=-1$.

\section{Exponent Estimation}

The exponents have been estimated using maximum likelihood estimation together with a discrete exponentially truncated power-law based on the method used in Ref. \cite{alstott2014}. The use of maximum likelihood estimation is motivated by the fact that we look at probability distributions. The estimated exponents for different lattice sizes~$L$ have been observed to converge algebraically to some limit value $\gamma_0$ as $L\rightarrow\infty$. This limit is extrapolated by fitting the estimated exponents~$\gamma(L)$ against $\gamma(L) = \gamma_0 + \gamma_s L^{\gamma_e}$ using non-linear least squares.

\section{Conclusion}

We have investigated the effect of flipping one edge in planar directed networks (diode networks) at the critical probability $p_c$ and found power-law behaviours for, both, the probability distribution of the differences in shortest path lengths and for the minimal enclosed areas between the old and new shortest paths. This implies that very small perturbations made to the shortest path can cause non-local changes in the system. We found that the number of divergent paths vanishes in the thermodynamic limit. The differences in shortest path lengths are thus expected to be even-valued (or zero) in the thermodynamic limit. We also give estimates for the exponents of the observed power-laws using maximum likelihood estimates based on a truncated power-law and extrapolation. The exponents are $\alpha = 1.36\pm 0.01$ for the differences in shortest path lengths and $\beta = 1.186\pm 0.001$ for the minimal enclosed areas. 

Finally, we found that the value we obtain for the exponent $\beta$ agrees within the error-bars with the exponent $\beta=1.16\pm0.03$ reported in Ref. \cite{fehr2011}. Both cases deal with a distribution that is the result of a small perturbation of a fractal path.  In both cases a new path emerges after the perturbation, creating an enclosed area that is distributed according to the same power law, suggesting that the resistor-diode network and watersheds might be related to each other.

%\acknowledgments


\begin{thebibliography}{00}

\bibitem{redner1981} S. Redner, \textit{Percolation and conduction in a random resistor-diode network}, J. Phys. A \textbf{14} L349 (1981).
\bibitem{redner1982a} S. Redner, \textit{Directed and diode percolation}, Phys. Rev. B \textbf{25} 3242 (1982).
\bibitem{redner1982b} S. Redner, \textit{Conductivity of random resistor-diode networks}, Phys. Rev. B \textbf{25} 5646 (1982).
\bibitem{broadbent1957} S.R. Broadbent and J.M. Hammersley,\textit{ Proc. Cambridge Philos. Soc.} \textbf{53}, 629 (1957).
\bibitem{noronha2018} A.W.T Noronha, A.A. Moreira, A.P. Vieira, H.J. Herrmann, J.S. Andrade Jr. and H.A. Carmona, \textit{Peroclation on isotropically directed lattice}, arXiv:1808.06644.
\bibitem{janssen2000} H.-K. Janssen and O. Stenull, \textit{Random resistor-diode networks and the crossover from isotropic to directed percolation}, Phys. Rev. E. \textbf{62} 3173 (2000).

\bibitem{havlin1984} S. Havlin and R. Nossal, \textit{Topological properties of percolation clusters}, J. Phys. A \textbf{17} L427 (1984).
\bibitem{grassberger1983} P. Grassberger, \textit{On the critical behavior of the general epidemic process and dynamical percolation}, Math. Biossci. \textbf{63} 157 (1983).
%\bibitem{alexandrowicz1980} Z. Alexandrowicz, \textit{Critically branched chains and percolation clusters}, Phys. Lett. A \textbf{80} 284 (1980).
\bibitem{pike1981} R. Pike and H.E. Stanley, \textit{Order propagation near the percolation threshold}, J. Phys. A \textbf{14} L169 (1981).
\bibitem{herrmann1984} H.J. Herrmann, D.C. Hong and H.E. Stanley, \textit{Backbone and elastic backbone of percolation clusters obtained by the new method of 'burning'}, J. Phys. A \textbf{17} L261 (1984).

\bibitem{grassberger1985} P. Grassberger, \textit{On the spreading of two-dimensional percolation}, J. Phys. A \textbf{18} L215 (1985).
\bibitem{herrmann1988} H.J. Herrmann and H.E. Stanley, \textit{The fractal dimension of the minimum path in two- and three-dimensional percolation}, J. Phys. A \textbf{21} L829 (1988).
\bibitem{zhou2012} Z. Zhou, J. Yang, Y. Deng and R.M. Ziff, \textit{Shortest-path fractal dimension for percolation in two and three dimensions}, Phys. Rev. E \textbf{86} 0611101 (2012).


\bibitem{havlin1985} S. Havlin, B. Trus, G.H. Weiss and D. Ben-Avraham, \textit{The chemical distance distribution in percolation clusters}, J. Phys. A \textbf{18} L247 (1985).
\bibitem{neumann1988} A. U. Neumann and S. Havlin,\textit{ Distributions and Moments of Structural Properties for Percolation Clusters}, J. Stat. Phys. \textbf{52} 203 (1988).
\bibitem{aharony1992} A. Aharony and A.B. Harris, \textit{Multifractal localization}, Physica A \textbf{191} 365 (1992).
\bibitem{hovi1997} J. P. Hovi and A. Aharony, \textit{Renormalization group calculation of distribution functions: Structural properties for percolation clusters}, Phys. Rev. E \textbf{56} 172 (1997).


\bibitem{zhou2012b} Z. Zhou, J. Yang, R.M. Ziff and Y. Deng, \textit{Crossover from isotropic to directed percolation}, Phys. Rev. E \textbf{86} 021102 (2012).

\bibitem{fehr2011} E. Fehr, D. Kadau, J. S. Andrade, and H. J. Herrmann, \textit{Impact of perturbations on watersheds}, Phys. Rev. Lett. \textbf{106} 048501 (2011).

\bibitem{alstott2014} J. Alstott, E. Bullmore, and D. Plenz, \textit{powerlaw: A Python Package for Analysis of Heavy-Tailed Distributions}, PLoS ONE \textbf{9} (2014).

	
\end{thebibliography}
\end{document}